\documentstyle[12pt,aps,prc,epsfig,preprint]{revtex}
\tightenlines
\begin{document}

\begin{titlepage}
\title{\vspace*{5mm}\bf
\Large Renormalization of the isovector  $\pi N$  amplitude in pionic atoms}
\vspace{4pt}

\author{  E.~Friedman and A.~Gal \\
{\it Racah Institute of Physics, The Hebrew University, Jerusalem 91904,
Israel\\}}

\vspace{4pt}
\maketitle

\begin{abstract}

The extraction of the isovector  $s$-wave $\pi N$  amplitude from pionic
atoms is studied with special emphasis on uncertainties and their 
dependence on the 
 {\it assumptions}  made regarding the neutron density
distributions in nuclei and on the size of the data base used . 
 Only `global' analyses of pionic-atom
data reveal a discrepancy between the extracted
isovector  $s$-wave $\pi N$  amplitude $b_1=-0.108\pm0.007~~ m_\pi^{-1}$ 
and its free $\pi N$ counterpart 
$b_1^{{\rm free}}=-0.0885^{+0.0010}_{-0.0021} ~~m_\pi ^{-1}$, where the
uncertainty in the neutron densities is included in the error analysis.
The role of `deeply bound' pionic atom states is discussed and the 
reason for failure
of these states to provide new information is explained.

$PACS$: 13.75.-n; 13.75.Gx; 25.80.Hp
\newline
\newline \newline
%\vspace{1cm}
Corresponding author: E. Friedman, \newline
Tel: +972 2 658 4667,
FAX: +972 2 658 6347, \newline
E mail: elifried@vms.huji.ac.il

\end{abstract}

\centerline{\today}
\end{titlepage}

\section{Introduction}
\label{sec:int}

In the past few years there has been renewed interest 
\cite{FGa98,KYa01,Fri02,Fri02a,FGa02,GGG02a,GGG02,SFG02,YHi03}
in the extraction of the isovector  $s$-wave 
$\pi N$  scattering amplitude near threshold
from pionic atom observables. Of particular
interest is the relationship between this in-medium
amplitude  and the corresponding
amplitude for the free $\pi N$ interaction at threshold. 
This renewed interest in pionic atoms in general, and in the $s$-wave
part of the potential in particular, stems from three recent
developments. The first one is the experimental observation of `deeply
bound' pionic atom  states in the (d,$^3$He) reaction
\cite{GGG02,SFG02,YHI96,GGK00}, the existence of which was 
predicted a decade earlier
\cite{FSo85,TYa88,THY89}. The second one is the very accurate measurements
of the shift and width of the 1$s$ level in pionic
hydrogen \cite{SBG01} and in pionic deuterium \cite{CEJ97,HKS98}
which leads to precise values for the $s$-wave
$\pi N$ scattering lengths (see also Ref. \cite{ELT02}).
The third development is the attempt to explain the `anomalous' 
$s$-wave repulsion
\cite{BFG97} in terms of a density dependence of the pion decay constant
\cite{Wei01}, or very recently by 
constructing the $\pi N$ amplitude near threshold within a systematic
chiral perturbation expansion \cite{KWe01}
and in particular imposing on it gauge invariance \cite{KKW02,KKW02a}.

The so-called anomalous repulsion of the $s$-wave pionic atom potential
is the empirical finding, from fits of optical potentials to pionic atom
data, that the strength of the repulsive $s$-wave
potential inside nuclei is typically twice as large as is expected on
the basis of the free $\pi N$ interaction. 
It is interesting to note that early indications for a very repulsive
isovector $s$-wave amplitude go as far back as 1978 \cite{BBF78}
when high precision data became available. Subsequent analyses \cite{GNO92}
also indicated such a repulsive amplitude.
A key point in this problem
are the uncertainties which are associated with the empirical determination
of the relevant potential parameters. This problem has been aggravated 
by recent works in which the `deeply bound' pionic atom states were
granted special status and unfounded implicit assumptions were made
on the pion-nucleus optical potential (e.g. Refs. \cite{GGG02a,SFG02}). 
The present
work addresses all these issues with the aim of obtaining reliable
parameter values for the $s$-wave part of the pion-nucleus optical
potential at threshold, together with realistic estimates of uncertainties.
The focus of this work is on the extraction of $b_1$, the in-medium
{\it isovector} $s$-wave $\pi N$ amplitude, from pionic atom data, 
using the conventional (density-independent) $\pi$-nucleus optical
potential. It is shown that no definitive conclusions can be reached
on whether or not $b_1$ is renormalized with respect to its free-space
value unless the whole bulk of pionic-atom data across the periodic table 
is used in the analysis. In particular, the deeply bound 1$s$ states
observed recently in the (d,$^3$He) reaction
 \cite{GGG02,SFG02} do not offer conclusive
evidence for a renormalization of $b_1$. That applies also to analyses
using the deeply bound states  together with `normal' 1$s$
states \cite{GGG02a}.

Section \ref{sect:potl} presents briefly the 
conventional pion-nucleus potential \cite{EEr66} where
the various parameters are defined.
Section \ref{sect:radii} is devoted to a discussion of the proton and
neutron density distributions which are an essential ingredient of
the pion-nucleus optical potential. The proton densities
obviously are obtained from the quite well known charge distributions
of nuclei but the neutron distributions are generally not known to a
sufficient accuracy. 
For this reason simple but physical parameterizations
of the neutron density distributions are introduced, in comparison with 
available data and with nuclear models.
The dependence of the isovector $s$-wave $\pi N$
amplitude on the neutron distributions was only marginally touched upon
in recent works \cite{Fri02a,GGG02a} and, therefore, we study  
in Section \ref{sect:results} in some
detail the dependence of this amplitude on both the root mean square  
(rms) radius of the distribution and,
to some extent, on the shape of the distribution. This dependence is
translated then into an additional uncertainty of the extracted 
parameter values.
Section \ref{sect:results} also presents the results of fits to the data,
starting with `global' fits to 100 and 120 data points 
and gradually reducing the extent of the data base in order to assess
the accuracy of the results, in comparison with other works using
`partial' fits to very limited sets of data \cite{GGG02a,SFG02,YHi03} 
that place special 
emphasis  on  the deeply bound 1$s$ states in
$^{115,119,123}$Sn and in $^{205}$Pb. 
The last subsection of Section \ref{sect:results} discusses the
Seki-Masutani linearization scheme \cite{SMa83} demonstrating the limits
of its applicability to pionic-atom studies.
In Section \ref{sect:deep} we
explain why the deeply bound pionic atom states generally have failed to
provide new information on the pion-nucleus interaction. A brief
summary and conclusions are presented in Sect. \ref{sect:summ}.

\section{The pion-nucleus potential} \label{sect:potl}

The pion-nucleus potential has been extensively discussed 
(e.g. Ref. \cite{BFG97}),
but for completeness a short summary follows.

The interaction of pions at threshold with the nucleus
is described by the Klein-Gordon  
 equation of the form:

\begin{equation}\label{equ:KG1}
\left[ \nabla^2  - 2{\mu}(B+V_{{\rm opt}} + V_c) + (V_c+B)^2\right] \psi = 0~~ ~~
(\hbar = c = 1)
\end{equation}
where $\mu$ is the pion-nucleus reduced mass,
$B$ is the complex binding energy
 and $V_c$ is the finite-size
Coulomb interaction of the pion with the nucleus, including
vacuum-polarization terms.
The optical potential $V_{\rm opt}$ used in the present work is of 
the canonical form due to  Ericson and Ericson \cite{EEr66},

\begin{equation} \label{equ:EE1}
2\mu V_{{\rm opt}}(r) = q(r) + \vec \nabla \cdot \alpha(r) \vec \nabla
\end{equation}
with
\begin{eqnarray} \label{equ:EE1s}
q(r) & = & -4\pi(1+\frac{\mu}{M})\{{\bar b_0}(r)[\rho_n(r)+\rho_p(r)]
  +b_1[\rho_n(r)-\rho_p(r)] \} \nonumber \\
 & &  -4\pi(1+\frac{\mu}{2M})4B_0\rho_n(r) \rho_p(r) ,
\end{eqnarray}

\begin{equation} \label{equ:LL2}
\alpha (r) = \frac{\alpha _1(r)}{1+\frac{1}{3} \xi \alpha _1(r)}
 + \alpha _2(r) ,
\end{equation}
\noindent
where

\begin{equation} \label{equ:alp1}
\alpha _1(r) = 4\pi (1+\frac{\mu}{M})^{-1} \{c_0[\rho _n(r)
  +\rho _p(r)] +  c_1[\rho _n(r)-\rho _p(r)] \} ,
\end{equation}

\begin{equation} \label{equ:alp2}
\alpha _2(r) = 4\pi (1+\frac{\mu}{2M})^{-1} 4C_0\rho _n(r) \rho _p(r).
\end{equation}

\noindent
In these expressions $\rho_n$ and $\rho_p$ are the neutron and proton density
distributions normalized to the number of neutrons $N$ and number
of protons $Z$, respectively, %$\mu$ is the pion-nucleus reduced mass
and $M$ is the mass of the nucleon,
 $q(r)$ is referred to
as the $s$-wave potential term and $\alpha(r)$ is referred to
as the $p$-wave potential term.
The function ${\bar b_0}(r)$ in Eq. (\ref{equ:EE1s}) 
is given in terms of the {\it local} Fermi
momentum $k_{\rm F}(r)$ corresponding to the isoscalar nucleon
density distribution:

\begin{equation} \label{equ:b0b}
{\bar b_0}(r) = b_0 - \frac{3}{2\pi}(b_0^2+2b_1^2)k_{\rm F}(r),
\end{equation}
where $b_0$ and $b_1$ are minus the pion-nucleon isoscalar
and isovector effective scattering lengths, respectively.
The quadratic terms in $b_0$ and $b_1$ represent double-scattering
modifications of $b_0$. In particular, the $b_1^2$ term represents
a sizable correction to the nearly vanishing linear $b_0$ term.
Similar modifications of $b_1$ \cite{KWe01,KKW02} were found by us
to be negligibly small.
The  coefficients $c_0$ and
$c_1$ in Eq. (\ref{equ:alp1}) are the pion-nucleon
isoscalar and isovector $p$-wave scattering volumes,
respectively. The parameters $B_0$ and $C_0$ 
in Eqs. (\ref{equ:EE1s}) and (\ref{equ:alp2})   
represent $s$-wave and $p$-wave
absorption, respectively, on pairs
of nucleons and as such they have imaginary parts.
Dispersive real parts are found
to play an important role in pionic atom potentials.
The terms with $4\rho _n \rho _p$ were originally written as
$(\rho _n+\rho _p)^2$, but the results hardly
depend on which form is used.
The parameter $\xi$ in Eq. (\ref{equ:LL2}) is the usual
Ericson-Ericson Lorentz-Lorenz (EELL) coefficient  \cite{EEr66}.
An additional relatively small term,
known as the `angle-transformation' term (see Eq.(24) of \cite{BFG97}),
is also included.

The pion-nucleus optical potential contains nine parameters
but only six parameters were varied in the present fits, namely, all the
parameters of the $s$-wave part of the potential and the empirical
complex $p$-wave absorption parameter $C_0$. The linear $p$-wave parameters
$c_0$ and $c_1$ were kept fixed at their respective free $\pi N$ values,
as discussed extensively in Ref. \cite{Fri02a}.
The EELL coefficient  
was held fixed at the value  $\xi$ = 1.
It was demonstrated in Ref. \cite{Fri02a} that the extracted $s$-wave
parameters are insensitive, with negligibly small variations, to sizable
variations of the $p$-wave parameters, {\it provided} a fit to the data
was made.
Note that the extracted parameters  $b_0$ and $b_1$ are some averages over
nuclear densities of in-medium quantities, and if they turn out to be
different from the corresponding free $\pi N$ values then one could
refer to a renormalization of the free space values in the nuclear medium.

\section{Nuclear density distributions}
\label{sect:radii}
It is evident from the previous section that the nuclear density
distributions $\rho_p$ and $\rho_n$ are an essential part of the 
pion-nucleus potential. 
Nevertheless, their effect in the present context of studying the
$s$-wave part of the potential and in particular the dependence of
the parameter $b_1$ on the choice of nuclear densities has been 
hardly addressed, and only briefly so in Refs. \cite{Fri02a,GGG02a}.
This point is studied in more detail in the present work.

Nuclear charge densities are known quite well \cite{FBH95} from studies of
elastic scattering of electrons and from finite-size effects in
muonic atoms. Density distributions for protons can then be obtained by
unfolding the finite size of the charge of the proton. This leads to
reliable proton densities,
particularly in the surface region, which is the relevant region for
producing strong-interaction effects in pionic atoms. The neutron distributions
are, however, generally not known to sufficient accuracy. 
A host of different methods have been
applied to the extraction of rms radii of neutron distributions in nuclei
\cite{BFG89}, including some studies using pionic atoms \cite{GNO92},
but the results are sometimes conflicting. Several nuclei have been studied
more than others, e.g. isotopes of Ca, Sn and Pb \cite{BFG89,SHi94,KFA99,CKH02}
but for others there is
no experimental information on neutron densities and one must then
rely on calculations based on various models.
To complicate things further we note that there is a long history of 
conflict between values of neutron rms radii derived from  
experiments using hadronic projectiles and neutron rms radii derived
from theoretical calculations. For that reason we have adopted a 
semi-phenomenological approach that covers a broad range of possible
neutron density distributions.

\begin{figure}
\epsfig{file=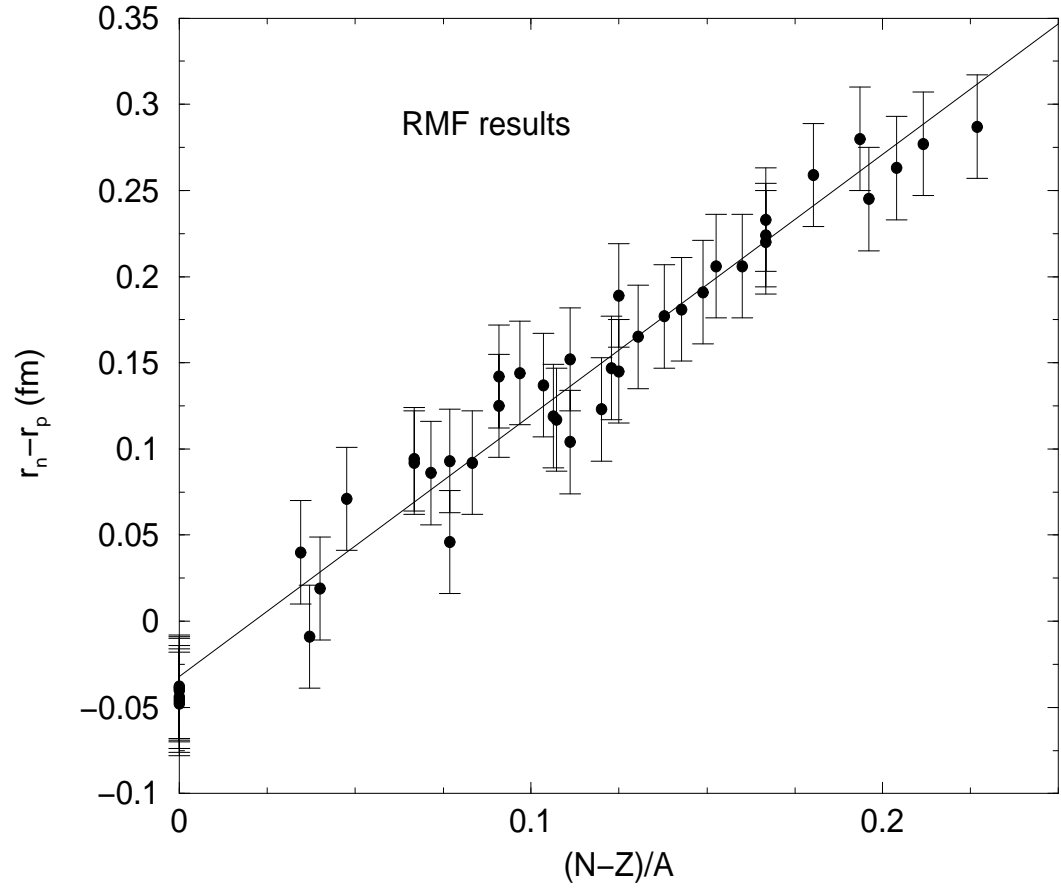, height=120mm,width=140mm}
%\vspace*{5mm}
\caption{Fit of a linear expression in the asymmetry parameter 
to RMF  values of $r_n-r_p$.}
\label{fig:RMF}
\end{figure}

It is shown below that
the feature of  neutron density distributions which is most effective
in  determining strong interaction level shifts and widths 
in pionic atoms is the radial extent, as represented, for
example, by $r_n$, the neutron density rms radius. Other features 
such as  the detailed 
shape of the distribution have only minor effect. For that reason we chose
the rms radius as the prime parameter in the present study. 
 Since $r_p$, the rms radius for the 
proton density distribution, is considered to be known, we focus attention
on values of the difference $r_n-r_p$. In the analysis of Ref. \cite{Fri02a} 
the nuclear densities were either of the single particle (SP) variety or
of the macroscopic type \cite{BFG97} where the values of $r_n-r_p$ were
taken from the relativistic mean field (RMF) results of Lalazissis et al.
\cite{LRR99} for each of the nuclei involved. Smaller values of  $r_n-r_p$
were also used in Ref. \cite{Fri02a}.
In the present work we vary, among other things, the values of $r_n-r_p$
and we therefore look for a simple 
parameterization that will be easy to relate to the RMF results.
Choosing 43 stable nuclei along the periodic table, 
the calculated RMF values \cite{LRR99}
of $r_n-r_p$  are  described very well by the expression

\begin{equation} \label{equ:RMF}
r_n-r_p = \alpha \frac{N-Z}{A} + \beta .
\end{equation}
Figure \ref{fig:RMF} shows the results of this linear fit in the
neutron-proton asymmetry $(N-Z)/A$, where we have
arbitrarily assigned an error of $\pm$0.03 fm to each value of $r_n-r_p$,
as  a representative value of discrepancies between calculated and
experimental charge rms radii. The RMF fit parameters obtained are
$\alpha$=1.51$\pm$0.07~fm, $\beta$=$-$0.03$\pm$0.01 fm.

An expression of the form (\ref{equ:RMF}) has also been used recently
in connection with antiprotonic atoms \cite{TJL01}, with parameter values
of $\alpha$=1.01$\pm$0.15 fm, $\beta$=$-$0.04$\pm$0.03 fm.
Note that antiprotonic atoms are sensitive to nuclear densities well
below 10\% of the central density and as such it is 
not very reliable to infer from these data quantities which relate to the bulk
of the density.  Nevertheless we consider these
results as an experimental indication on values of $r_n-r_p$,
particularly as they have been used \cite{GGG02a} in analyses of pionic atoms.
In the present work the neutron rms radii were then varied using 
Eq.(\ref{equ:RMF}) with $\beta$=$-0.035$ fm and varying the parameter
$\alpha$ between 0.3 and 2.7 fm. This way we have covered a very wide range
of possible values of neutron rms radii.

In order to allow for possible differences in the shape of the neutron
distribution, the `skin' and the `halo' forms of Ref. \cite{TJL01} were
used, as well as an average between the two. Assuming a two-parameter
Fermi distribution both for the proton (unfolded from the charge distribution)
and for the neutron density distributions 

\begin{equation}
\rho_{n,p}(r)  = \frac{\rho_{0n,0p}}{1+{\rm exp}((r-R_{n,p})/a_{n,p})},
\end{equation}
then for each value of $r_n-r_p$ in the skin form  
the same diffuseness parameter for the protons and the neutrons 
$a_n=a_p$ was used and the $R_n$ parameter was determined from the 
rms radius $r_n$. 
In the halo form
the same radius parameter $R_n=R_p$ was assumed and $a_n^{\rm h}$ was
determined from $r_n$. In the `average'
option the diffuseness parameter was set to be the average of the 
above two diffuseness parameters $a_n^{{\rm ave}}=(a_p+a_n^{\rm h})/2$
and the radius parameter $R_n$ was determined from the 
rms radius $r_n$. In this
way we have used three shapes of the neutron distribution for each value
of its rms radius all along the periodic table.

\section{Results and discussion} \label{sect:results}
\subsection{Dependence on neutron distribution}
For the study of the dependence on the neutron density distribution 
using the parameterization of Eq. (\ref{equ:RMF}) above we first chose
a data base of 100 points, from $^{20}$Ne to $^{238}$U, 
including the recently published results  for 
the deeply bound 1$s$ states in $^{115,119,123}$Sn 
\cite{SFG02} and those for the
1$s$ and 2$p$ states in $^{205}$Pb \cite{GGG02}.
Ne was chosen as the lightest element because it is the 
first element in the periodic table for which the RMF results \cite{LRR99}
apply and because we use the two-parameter Fermi parameterization for the
densities. 
Fits to the data were made by varying six of the parameters of
the potential, as discussed is Sect. \ref{sect:potl}.

\begin{figure}
\epsfig{file=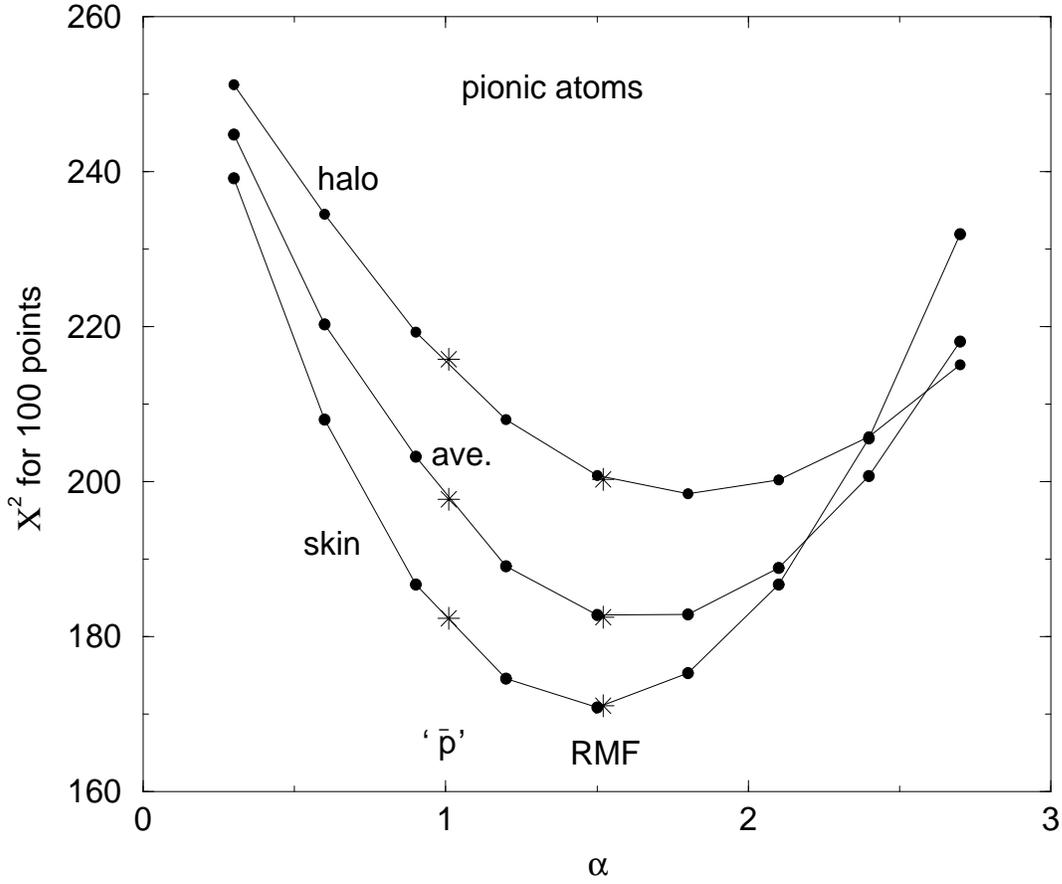, height=120mm,width=140mm}
%\vspace*{5mm}
\caption{Values of $\chi ^2$ for 100 data points from $^{20}$Ne to $^{238}$U
as function of the slope parameter $\alpha $ in Eq. (\ref{equ:RMF})
for three shapes of densities (see text). Points corresponding to `$\bar p$'
and `RMF' values for $\alpha $ are marked with  asterisks.}
\label{fig:chi2}
\end{figure}

Figure \ref{fig:chi2} shows $\chi ^2$ values for the three types of neutron
density, namely, skin, halo and average,
as a function of  $\alpha$, the coefficient of the asymmetry 
parameter $(N-Z)/A$ which
determines the value of $r_n-r_p$, 
see Eq. (\ref{equ:RMF}). Also marked by asterisks
are the points  associated with the particular values of 
$\alpha$=1.01 fm and $\alpha$=1.51 fm, which refer to the `$\bar p$' and the
RMF prescriptions, respectively.
It is easy to conclude from this figure that 
in the context of global fits to pionic atom data
the RMF radii and the skin
shape are the favoured option. Note that the minimum of $\chi ^2$ 
implies $\chi ^2 $/F, the $\chi ^2$  per degree of freedom, of 1.8, 
which represents a very good fit.
Increasing the data base to 120 points by adding 20 points from $^{12}$C
to $^{19}$F and repeating the fits, the overall picture remains 
essentially the same, but the minima
 become less sharp and imply  slightly larger
values of $\chi ^2 $/F.
For these additional nuclei 
(except $^{19}$F) a modified harmonic oscillator form \cite{FBH95}
was used for the densities.

The above results may be interpreted as indicating that on the average
the difference $r_n-r_p$ is described rather well by the RMF 
model ($\alpha$=1.5 fm).
However, when other experimental data had been analyzed in terms
of the difference between rms radii \cite{BFG89,SHi94,KFA99,CKH02}, 
quite often smaller values were obtained for $r_n-r_p$, corresponding
to $\alpha$ between 1 and 1.5 in Eq. (\ref{equ:RMF}).
Note that Brueckner-Hartree-Fock calculations generally yield values
of $r_n-r_p$ corresponding to $\alpha \sim $ 1.0 \cite{PRL97}.
We therefore conclude that on the average, rms radii of neutron density
distributions are between the values inferred from analyses of 
antiprotonic atoms  ($\alpha$ =1.0) \cite{TJL01} and the predictions of RMF 
calculations \cite{LRR99}.

\begin{figure}
\epsfig{file=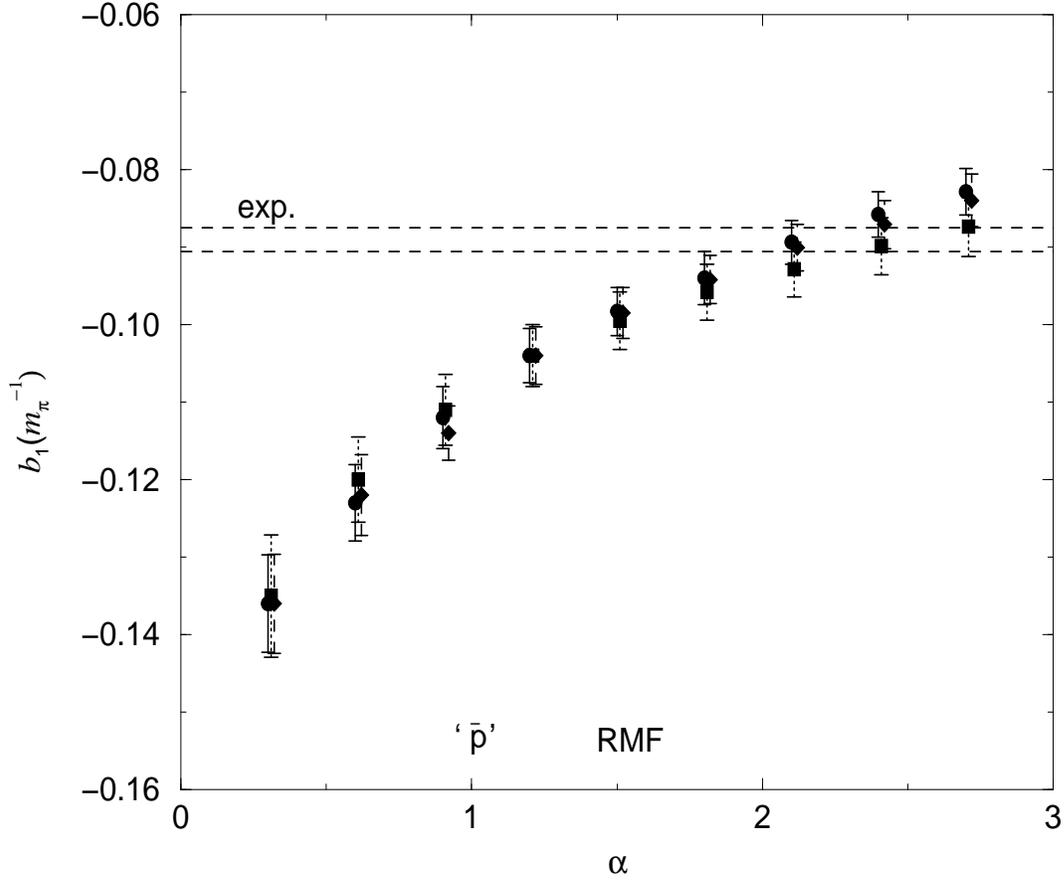, height=120mm,width=140mm}
%\vspace*{5mm}
\caption{Values of $b_1$ from fits to 100 data points from 
$^{20}$Ne to $^{238}$U
as function of the slope parameter $\alpha $ 
 for three shapes of densities
(see Fig. \ref{fig:chi2}). 
Also shown is the experimental value  {\protect \cite{SBG01}} 
for $b_1^{\rm free}$ .} 
\label{fig:b1radii}
\end{figure}

Turning to the question of the dependence of extracted parameters of the
pion-nucleus optical potential on the neutron distribution used in the
analysis,
Fig. \ref{fig:b1radii} shows the values of $b_1$ which correspond to the 
 $\chi ^2$ results displayed in Fig. \ref{fig:chi2}.
It is evident that there is a fairly strong dependence
of the extracted $b_1$ on  the prescription chosen for 
$r_n-r_p$, whereas the dependence
on the shape of the density distribution, as indicated by the scatter of
points for each value of $\alpha$, is small.
A similar plot of the results of fits to the 120 data points mentioned
above is almost indistinguishable from the present figure.
Adopting an average value of $\alpha$= 1.25$\pm$0.25 fm as 
discussed above, one can
estimate the additional uncertainty  of the extracted potential
parameters, due to neutron densities,
 with the help of plots like the one shown in Fig. \ref{fig:b1radii}.

\subsection{Reduced data bases}
Several recent publications \cite{GGG02a,SFG02} have claimed exceedingly
small uncertainties for the parameter $b_1$ when it is extracted from
data bases which are very small  in comparison with the 100-120 points of 
data used above. Such
claims are based on adopting  several linearizations of terms
in the optical potential Eq. (\ref{equ:EE1s}), notably the 
Seki-Masutani  correlation \cite{SMa83},
 which are at variance with
the results of global fits to large scale data bases,
see Refs. \cite{Fri02,FGa02} and the following subsection.
We have, therefore, repeated the analysis using progressively reduced data
bases in order to examine {\it in a consistent manner} the dependence of the
uncertainties on the data,  including also  
the uncertainty in the neutron density ditribution. The results of this
survey are summarized in Tables \ref{tab:res1} and \ref{tab:res2}. 
The results in
Table \ref{tab:res1} are based on the use of the `RMF skin' neutron densities
and the results in Table \ref{tab:res2} are based on the use of
the `$\bar p$ skin' densities.
The `global 2' and `global 3' data sets are the ones mentioned in the
previous subsection (`global 1' was used in Ref. \cite{Fri02a,FGa02}).
Three `partial' data sets are included in the tables, as follows:
(i)1$s$ states in light pionic atoms from $^{12}$C to $^{28}$Si;
(ii)1$s$ states in light $N=Z$ pionic atoms plus the deeply bound
1$s$ states in $^{115,119,123}$Sn \cite{SFG02}  
and $^{205}$Pb \cite{GGG02}; and (iii)only the deeply
bound states of (ii). Within each table it is seen that the parameter
values are consistent with each other as the size of the data base is
changed, but the uncertainties increase as the data base becomes
smaller.  Note  that Re$B_0$ is different from
zero, within errors, only for the global fits. 
Claims based on `partial' data sets that no Re$B_0$ is required to fit the
data therefore do not hold for pionic atoms in general.
In Tables \ref{tab:res1} and \ref{tab:res2}, for the fit to only deeply bound 
states, we have set Re$B_0$=0 since
otherwise the convergence proved unreliable with so few data points.

Turning to the role played by the neutron density distributions,
 it is evident from Tables \ref{tab:res1} and \ref{tab:res2}
that the dependence of the extracted
parameters on the rms radius of the neutron density distribution must be
taken into account in  the evaluation of the final results.
Table \ref{tab:res3} summarizes the final values of $b_1$, 
 for $\alpha$=1.25$\pm$0.25~fm  (Eq. (\ref{equ:RMF})) as
discussed above. Since one deals here with an {\it average} behaviour of
$r_n-r_p$ along the periodic table, the effects of the uncertainty
$\Delta \alpha = \pm 0.25$ fm are included in quadrature.  
Again all the different
results for $b_1$ are consistent with each other, but
only those due to  global
fits to 100-120 data points may be regarded as significantly different from
the free $\pi N$ value \cite{SBG01} of
$b_1^{{\rm free}}=-0.0885^{+0.0010}_{-0.0021} ~~m_\pi ^{-1}$. Claiming such
disagreement  by analysing  reduced data sets can only be described
as unfounded.

\subsection{The Seki-Masutani linearization}

Having demonstrated the dependence of the uncertainties of the derived 
values of $b_1$ on the size of the data base, we turn now to another problem
which has affected recent derivations of this parameter 
\cite{GGG02a,GGG02,SFG02,YHi03},
namely, the use of the Seki-Masutani (SM) linearization \cite{SMa83}. 
In the SM scheme the Re$B_0$ term which is quadratic in the densities
is linearized with the help of an `effective' density such that the two
terms with $b_0$ and Re$B_0$ are lumped together into a single term
$b_0^* \rho(r)$.
The existence
of correlations between the parameters $b_0$ and Re$B_0$ has been known
for a long time \cite{SMa83} and it can be seen also in Tables \ref{tab:res1}
and \ref{tab:res2}. However, the same tables show also that when large data
bases are being used, then the value of each one of these two 
parameters is reasonably well determined,
although not to a very good accuracy. In order to explicitly show the 
consequences of using the SM linearization we have repeated the fits reported in 
Table \ref{tab:res1}  with Re$B_0$ set to zero. The results are summarized
in Table \ref{tab:res1a} where it is seen that for the large data bases the
values of $\chi ^2$ have increased significantly. 
Note that the `natural' unit
for measuring these increases is $\chi ^2 $/F, the  $\chi ^2$ per degree 
of freedom. For the smaller data sets the increases are not statistically
significant, in full agreement with the uncertainties for Re$B_0$ quoted
in   Tables \ref{tab:res1} and \ref{tab:res2}. 
It is concluded that  the SM correlation is valid, as may
be expected for any non-linear problem over some range of parameters,
but when the quality of fit 
$\chi ^2$ criteria are taken into consideration then
the use of the SM linearization is unjustified and
in the case of large data bases it biases the results.

Another interesting conclusion obtained from Table \ref{tab:res1a} is
that when Re$B_0$ is held fixed 
the values of $b_1$ and their uncertainties  hardly change compared
to the unrestricted analysis. The values of $b_0$ naturally change in order to
compensate for the corresponding changes in Re$B_0$
 but the uncertainties of $b_0$  go down by
almost an order of magnitude, which is  unphysical. 
Since the derivation of $b_1$ by Geissel et al. \cite{GGG02a} is
based on ``$b^*_0-b_1$ constraints" ($b^*_0$ closely related to our $b_0$),
we conclude that
the uncertainties of $b_1$ obtained in such limited analysis 
of deeply bound states,  which is 
based on the SM linearization, are bound to be  
 artificially small.

\section{The role of deeply bound states}
\label{sect:deep}
In order to understand the role played by the deeply bound states in
pionic atoms of $^{115,119,123}$Sn and of $^{205}$Pb in global
fits, we compare values of $\chi ^2 $/N, the total $\chi ^2$
per point obtained in the fits, with the corresponding values for the
contribution of the
deeply bound states. For the `global 2' fit,
using RMF-skin densities (Table \ref{tab:res1}),
 we have $\chi ^2 $/N=2.0
whereas the contribution of the deeply bound states  
amounts to 1.0 per point for these states.
For the `global 3' fit the numbers are 1.7 and 0.8, respectively.
Removing the deeply bound states from the `global 2' data and repeating
the fit, we find $\chi ^2 $/N=2.0 and then using the resulting parameters
to predict the shifts and widths of the deeply bound states, we get
$\chi ^2 $/N=1.2 for these states. Therefore in all three examples 
the $\chi ^2 $/N
for the deeply bound states are smaller than the average, signifying that
the deeply bound states do not depart from the  picture obtained for normal
pionic atoms along the periodic table. Limiting the discussion to only
1$s$ states over the periodic table and confronting
in Table \ref{tab:res3} the results for light nuclei only with
the results for light $N=Z$ nuclei plus the deeply bound states 
(the latter subset was advocated in Ref.\cite{GGG02a}), again
it is concluded that the deeply bound states do not provide new information
on the isovector $s$-wave amplitude $b_1$.
This observation is not entirely new and it was shown already in \cite{FGa98}
that the deeply bound states in $^{207}$Pb were in line with the
expectations based on normal pionic atom data.
A possible explanation was also indicated in Ref. \cite{FGa98}, namely,
``the same mechanism which causes the deeply bound states to be narrow also
masks the deep interior of nuclei where new effects could possibly 
be observed".

\begin{figure}
\epsfig{file=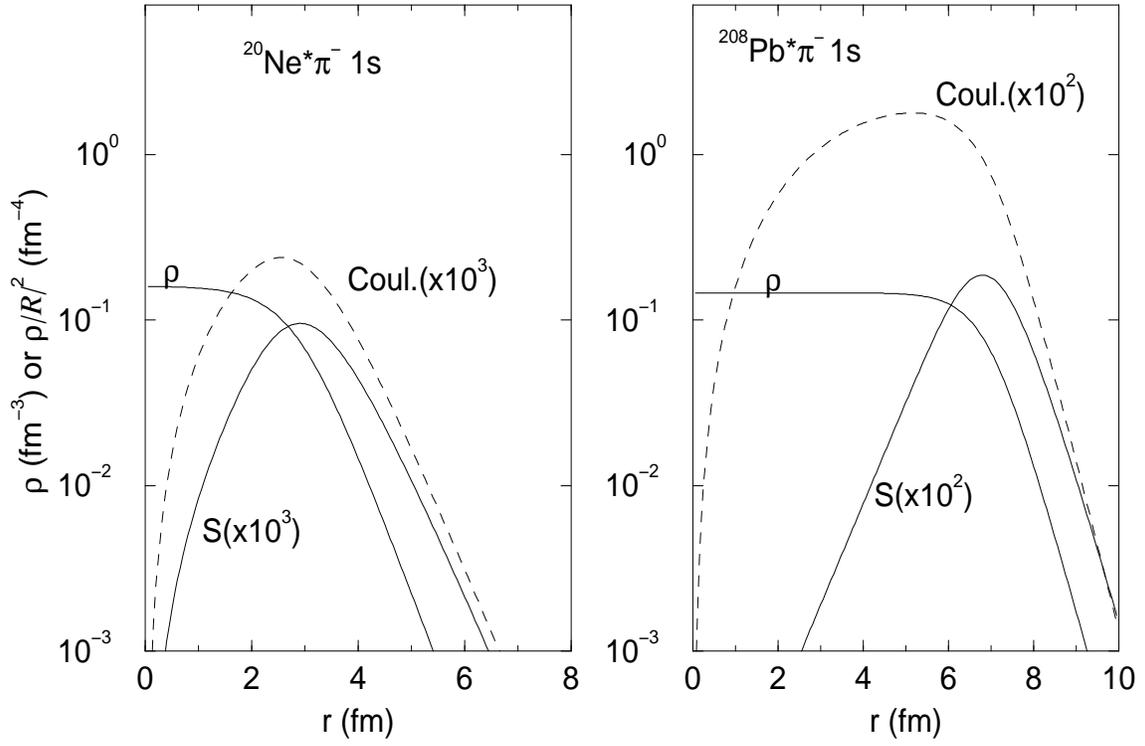, height=100mm,width=150mm}
%\vspace*{5mm}
\caption{Nuclear densities and radial densities for pionic $1s$ states 
times the nuclear density: dashed for (finite-size) Coulomb potential, 
solid curves (S) for Coulomb plus
strong interaction.}
\label{fig:overlap}
\end{figure}

The fact that the deeply bound pionic atom states have failed to provide
new information on the pion-nucleus interaction can be understood from simple
arguments of overlap between the atomic wavefunction and the nucleus
\cite{FGa02}. Figure \ref{fig:overlap}  displays absolute values squared of the atomic radial
wavefunctions multiplied by the nuclear density for a normal 1$s$ state
(in $^{20}$Ne) and for a deeply bound 1$s$ state (in $^{208}$Pb).
It is seen that the Coulomb wavefunction would have
indeed penetrated deeply into the heavy Pb nucleus, but due to the strong
interaction it is repelled such that its overlap with
the nucleus is sufficiently small to make the width of the state
relatively narrow and thus making the state observable. In fact, the
deeply bound 1$s$ wavefunction does not overlap with inner regions of the
nucleus more so than a normal 1$s$ wavefunction does. This shows clearly
why  deeply bound states do not play any special role
in the determination of pionic atom potentials. 
Expanding the statement made in Ref. \cite{FGa98} we
note that the same mechanism which causes the deeply bound
states to be narrow and observable, namely, the strong 
repulsion \cite{FSo85,TYa88} of
the wavefunction out of the nucleus, also masks the nuclear interior such
that the penetration of the deeply bound pionic atom wavefunction is not
dramatically enhanced compared to the normal states.

\section{Summary and conclusions}
\label{sect:summ}
In this work we have presented a comprehensive analysis of uncertainties
for the parameters of the pion-nucleus optical potential which are extracted
from experimental strong interaction effects in pionic atoms. Emphasis
was placed on the parameter $b_1$, the isovector $s$-wave $\pi N$ amplitude,
which has been discussed widely in recent years in connection with
possible medium modifications that could also explain the long-standing
`anomalous' repulsion. In addition to errors which are inherent in the
$\chi ^2$ fit procedure, we have examined errors which are introduced
because of the insufficient knowledge of neutron density distributions in
nuclei, both in shape and in radial extent. The uncertainties involved
in the use of partial data sets were compared with those obtained
from global analyses of large amount of data covering all the periodic table.
It was shown that only in the latter case the uncertainties are sufficiently
small that a discrepancy between  
the value $b_1=-0.108\pm0.007~~m_\pi ^{-1}$ derived 
 from pionic atoms and the corresponding
value for the free $\pi N$ interaction can be established.
In particular, the deeply bound states do not provide sufficiently extensive
and varied data base in order to constrain $b_1$ to the required precision.
The failure of the deeply bound states to provide 
meaningful new information on the
pion-nucleus interaction has been demonstrated and a simple explanation 
in terms of overlap arguments was offered.

Finally, we mention Weise's suggestion for a density dependence of $b_1$ 
motivated by a partial restoration of chiral symmetry in
dense matter \cite{Wei01}.
Its application was shown \cite{Fri02,Fri02a} to remove most of the anomaly
when applied to large data sets. Updating the fit termed `W' in \cite{Fri02a}
 for the present analysis of the 120 data points and including also the
uncertainties due to the neutron densities, we find 
the following low-density limit values:
$b_0=-0.009\pm0.007 ~~m_\pi ^{-1}$ and $b_1=-0.088\pm0.005~~m_\pi ^{-1}$,
in full agreement with the free $\pi N$ values \cite{SBG01} of 
$b_0^{{\rm free}}=-0.0001^{+0.0009}_{-0.0021} ~~m_\pi ^{-1}$ and
$b_1^{{\rm free}}=-0.0885^{+0.0010}_{-0.0021} ~~m_\pi ^{-1}$.
The parameter Re$B_0$ assumes then the value $-0.02\pm0.03~~ m_\pi ^{-4}$,
which is most acceptable (see Ref. \cite{Fri02}).
Although successful on a phenomenological level, this approach has
been deemed controversial. For example, Chanfray et al. \cite{CEO02}
have recently discussed the additional medium effect of a $\sigma $
field on the renormalization of the pion field in dense matter, finding
a considerably smaller renormalization effect than that suggested by
Weise. On the other hand, Kolomeitsev et al. \cite{KKW02} have argued 
that Weise's fairly substantial renormalization effect can be derived
by considering the energy dependence of $V_{{\rm opt}}$
within a systematic chiral perturbation expansion and enforcing
consistently gauge invariance. This latter suggestion will be addressed
in a forthcoming paper.

\vspace{5mm}
We wish to acknowledge the assistance of G. Yaari in the calculation
of nuclear densities.
This research was partially supported by the Israel Science Foundation.

\begin{table}
\caption{Parameter values from fits to various sets of pionic atom 
data based on RMF neutron radii.
The linear $p$-wave parameters were held fixed at $c_0$=0.22$~m_\pi^{-3}$,
$c_1$=0.18$~m_\pi^{-3}$ and $\xi$=1.}
\label{tab:res1}
\begin{tabular}{lccccc}
%\hline
data & `global 2'& `global 3'& light $N=Z$ & light $N=Z$ & `deep'\\
 &$^{12}$C to $^{238}$U&$^{20}$Ne to $^{238}$U& + light $N>Z$&
+`deep' & \\
 & & & 1$s$ only&1$s$ only & 1$s$ only\\
\hline
points       & 120 & 100 & 22 & 20 & 8\\
 $\chi ^2$ &237& 171& 54& 35& 1.6 \\
  $\chi ^2 $/F  & 2.1 & 1.8 & 3.0 & 2.2 & 0.3\\
$b_0 (m_\pi ^{-1})$ &~~0.000$\pm$0.006 &$-$0.001$\pm$0.007&
$-$0.009$\pm$0.017 &$-$0.016$\pm$0.013 & $-$0.007$\pm$0.016\\
$b_1 (m_\pi ^{-1})$ &$-$0.101$\pm$0.003 &$-$0.098$\pm$0.003&
$-$0.095$\pm$0.013 &$-$0.094$\pm$0.007 & $-$0.117$\pm$0.030 \\
Re$B_0(m_\pi ^{-4})$&$-$0.085$\pm$0.030&$-$0.082$\pm$0.030&
$-$0.048$\pm$0.072&$-$0.017$\pm$0.060 & 0 (fixed)\\
Im$B_0(m_\pi ^{-4})$&~~0.049$\pm$0.002&~~0.052$\pm$0.002&
~~0.049$\pm$0.002&~~0.051$\pm$0.002 & ~~0.058$\pm$0.007 \\
Re$C_0(m_\pi ^{-6})$&$-$0.011$\pm$0.008&$-$0.017$\pm$0.009&$-$0.011 (fixed)&
$-$0.011 (fixed)&$-$0.011 (fixed)\\
Im$C_0(m_\pi ^{-6})$&~~0.063$\pm$0.003&~~0.060$\pm$0.003&~~0.063 (fixed)&
~~0.063 (fixed)&~~0.063 (fixed)\\
\end{tabular}

`deep' refers to deeply bound 1$s$ states in $^{115,119,123}$Sn and
$^{205}$Pb.
\end{table}

\begin{table}
\caption{Parameter values from fits to various sets of pionic atom 
data based on `$\bar{p}$' neutron radii.
The linear $p$-wave parameters were held fixed at $c_0$=0.22$~m_\pi^{-3}$,
$c_1$=0.18$~m_\pi^{-3}$ and $\xi$=1.} 
\label{tab:res2}
\begin{tabular}{lccccc}
%\hline
data & `global 2'& `global 3'& light $N=Z$ & light $N=Z$ & `deep'\\
 &$^{12}$C to $^{238}$U&$^{20}$Ne to $^{238}$U& + light $N>Z$&
+`deep' & \\
 & & & 1$s$ only&1$s$ only & 1$s$ only\\
\hline
points       & 120 & 100 & 22 & 20 & 8\\
 $\chi ^2 $ & 270 &182 & 65 & 35 & 2.1 \\
  $\chi ^2 $/F  & 2.4 & 1.9 & 3.6 & 2.2 & 0.4\\
$b_0 (m_\pi ^{-1})$ &~~0.009$\pm$0.006 &~~0.001$\pm$0.007&
$-$0.015$\pm$0.017 &$-$0.011$\pm$0.015 & ~~0.003$\pm$0.022\\
$b_1 (m_\pi ^{-1})$ &$-$0.114$\pm$0.004 &$-$0.109$\pm$0.004&
$-$0.103$\pm$0.014 &$-$0.114$\pm$0.009 & $-$0.143$\pm$0.037 \\
Re$B_0(m_\pi ^{-4})$&$-$0.112$\pm$0.035&$-$0.085$\pm$0.035&
$-$0.012$\pm$0.072&$-$0.017$\pm$0.072 & 0 (fixed) \\
Im$B_0(m_\pi ^{-4})$&~~0.049$\pm$0.002&~~0.054$\pm$0.003&
~~0.048$\pm$0.002&~~0.051$\pm$0.002 & ~~0.065$\pm$0.008 \\
Re$C_0(m_\pi ^{-6})$&$-$0.010$\pm$0.008&$-$0.010$\pm$0.009&$-$0.010 (fixed)&
$-$0.010 (fixed)&$-$0.010 (fixed)\\
Im$C_0(m_\pi ^{-6})$&~~0.063$\pm$0.003&~~0.057$\pm$0.004&~~0.063 (fixed)&
~~0.063 (fixed)&~~0.063 (fixed)\\
\end{tabular}

`deep' refers to deeply bound 1$s$ states in $^{115,119,123}$Sn and
$^{205}$Pb.
\end{table}

\begin{table}
\caption{Average values of $b_1$ with added uncertainties due to
neutron distributions.  The free pion-nucleon value {\protect \cite{SBG01}} is
$b_1^{\rm free}=-0.0885^{+0.0010}_{-0.0021}~m_\pi ^{-1}$.}
\label{tab:res3}
\begin{tabular}{lccccc}
%\hline
data & `global 2'& `global 3'& light $N=Z$ & light $N=Z$ & `deep'\\
 &$^{12}$C to $^{238}$U&$^{20}$Ne to $^{238}$U& + light $N>Z$&
+`deep' & \\
 & & & 1$s$ only&1$s$ only & 1$s$ only\\
\hline
points       & 120 & 100 & 22 & 20 & 8\\
$b_1 (m_\pi ^{-1})$ &$-$0.108$\pm$0.007 &$-$0.104$\pm$0.006&
$-$0.099$\pm$0.014 &$-$0.104$\pm$0.013 & $-$0.130$\pm$0.036 \\
\end{tabular}

`deep' refers to deeply bound 1$s$ states in $^{115,119,123}$Sn and
$^{205}$Pb.
\end{table}

\begin{table}
\caption{Same as Table \ref{tab:res1} but with Re$B_0$=0 (fixed) showing
the deterioration in the quality of fits as measured by
 $\frac{\Delta \chi ^2}{\chi ^2 /{\rm F}}$, the increase of $\chi ^2$ in
units of $\chi ^2 $ per degree of freedom.}
\label{tab:res1a}
\begin{tabular}{lcccc}
%\hline
data & `global 2'& `global 3'& light $N=Z$ & light $N=Z$ \\
 &$^{12}$C to $^{238}$U&$^{20}$Ne to $^{238}$U& + light $N>Z$&
+`deep'  \\
 & & & 1$s$ only&1$s$ only \\
\hline
points       & 120 & 100 & 22 & 20 \\
 $\chi ^2 $ & 259 &190 & 56 & 35  \\
  $\chi ^2 $/F  & 2.3 & 2.0 & 2.9 & 2.1 \\
 $\frac{\Delta \chi ^2}{\chi ^2 /{\rm F}}$& 9.6 &10.6 & 0.7 & 0 \\
$b_0 (m_\pi ^{-1})$ &$-$0.018$\pm$0.001 &$-$0.019$\pm$0.001&
$-$0.020$\pm$0.003 &$-$0.020$\pm$0.002 \\
$b_1 (m_\pi ^{-1})$ &$-$0.102$\pm$0.003 &$-$0.099$\pm$0.003&
$-$0.093$\pm$0.012 &$-$0.094$\pm$0.007  \\
%Re$B_0(m_\pi ^{-4})$&$-$0.112$\pm$0.035&$-$0.085$\pm$0.035&
%$-$0.012$\pm$0.072&$-$0.017$\pm$0.072  \\
Im$B_0(m_\pi ^{-4})$&~~0.048$\pm$0.002&~~0.051$\pm$0.002&
~~0.048$\pm$0.002&~~0.050$\pm$0.002  \\
\end{tabular}

`deep' refers to deeply bound 1$s$ states in $^{115,119,123}$Sn and
$^{205}$Pb.

\end{table}

\end{document}